\documentclass{article}
\usepackage{cite}
\usepackage[cmex10]{amsmath}
\usepackage{algorithmic, algorithm}
\usepackage{array}
\usepackage{bm}
\usepackage[tight,footnotesize]{subfigure}
\usepackage{url}
\usepackage{bbm}
\usepackage{amsthm}
\usepackage{amsmath}
\usepackage{mathtools}
\usepackage{amssymb}
\usepackage{mathrsfs} 
\usepackage{graphicx}
\usepackage{color}
\usepackage{dsfont}
\usepackage{siunitx}

\newcommand{\twonorm}[1]{\left\|#1\right\|_{2}}
\newcommand{\infnorm}[1]{\left\|#1\right\|_{\ell_\infty}}

\newcommand{\abs}[1]{|#1|}

\newcommand{\mbf}[1]{\mathbf{#1}}

\newcommand{\rbf}{\mathbf{r}}

\newcommand{\ybf}{\mathbf{y}}
\newcommand{\sbf}{\mathbf{s}}
\newcommand{\xibf}{\boldsymbol{\xi}}

\newcommand{\Fbf}{\mathbf{F}}
\newcommand{\Pbf}{\mathbf{P}}

\newcommand{\Cibf}{\mathbf{C}_i}

\newcommand{\lbr}{\left\lbrace}
\newcommand{\rbr}{\right\rbrace}

\newcommand{\R}{\mathbb{R}}
\newcommand{\C}{\mathbb{C}}

\newcommand{\cj}{{ \rm j}}
\newcommand{\cT}{^{\rm H}}

\newcommand{\Fo}{\mathcal{F}}

\def\dint{\;\mathrm{d}}

\newtheorem{theorem}{Theorem}

\title{Accelerated Wirtinger Flow for \\Multiplexed Fourier Ptychographic Microscopy}

\author{Emrah Bostan$^{\ast}$, 
Mahdi Soltanolkotabi$^{\dagger}$, 
David Ren$^{\ast}$, and 
Laura Waller\thanks{\noindent Department of Electrical Engineering and Computer Sciences,
University of California, Berkeley \newline
$^{\dagger}$ Ming Hsieh Department of Electrical Engineering, University of Southern California \newline
E. Bostan's research is supported by the Swiss National Science
Foundation (SNSF) under grant P2ELP2 172278. 
M. Soltanolkotabi's research is supported by the Air Force Office of Scientific Research under
award number FA9550-18-1-0078.} }
\date{}
\begin{document}
\maketitle

\begin{abstract}
Fourier ptychographic microscopy enables gigapixel-scale imaging, with both
large field-of-view and high resolution. Using a set of low-resolution images
that are recorded under varying illumination angles, the goal is to
computationally reconstruct high-resolution phase and amplitude images. To
increase temporal resolution, one may use multiplexed measurements where the
sample is illuminated simultaneously from a subset of the angles. In this
paper, we develop an algorithm for Fourier ptychographic microscopy with such
multiplexed illumination. Specifically, we consider gradient descent type
updates and propose an analytical step size that ensures the convergence of the
iterates to a stationary point.  Furthermore, we propose an accelerated version
of our algorithm (with the same step size) which significantly improves the
convergence speed. We demonstrate that the practical performance of our
algorithm is identical to the case where the step size is manually tuned.
Finally, we apply our parameter-free approach to real data and validate its
applicability. 
\end{abstract}


\section{Introduction}
\label{Sec:Intro}

Fourier Ptychographic Microscopy (FPM) is a computational imaging technique
that---guided by synthetic aperture principles---generates gigapixel images
with both wide field-of-view and high resolution~\cite{Zheng.etal2013}.  The
system is  realized by replacing a microscope's illumination unit with a
light-emitting diode (LED) array. As LEDs illuminate the sample from different
angles, the camera captures multiple~\textit{intensity} images of different
spatial frequency bands of the sample, without moving parts. Based on a
nonlinear inverse problem, which is a type of phase
retrieval~\cite{Fienup.1982}, the low-resolution measurements are used to
computationally generate a high-resolution image of the sample in both
amplitude and phase.

FPM has been established as a viable tool for bioimaging
applications~\cite{Horstmeyer.etal2015,Tian.etal2015,Tian.Waller2015,
Chung.etal2015}, including \textit{live cell imaging}~\cite{Tian.etal2015} for
studies such as stem cell development and drug
discovery~\cite{Marrison.etal2013}.  Sequential data collection
(\textit{i.e.}~one image collected per LED) has limited temporal resolution,
preventing dynamic imaging. This has been addressed by \textit{multiplexed
coded illumination},~where a random subset of LEDs are turned on at the same
time to reduce the total number of images that need to be
taken~\cite{Tian.etal2014}. However, the multiplexed information must then be
decoupled, which can make the reconstruction less robust to noise and
model-mismatch~\cite{Ren.etal2017}. This puts emphasis on the stability (such
as convergence and sensitivity to hyper-parameters) of reconstruction
algorithms for FPM, especially when multiplexed illumination is used. 

Several algorithms have been proposed for solving the phase retrieval problem
within the context of FPM~\cite{Yeh.etal2015}. Existing reconstructions have
noticeably capitalized on gradient descent type methods (and their projected
variants) for the multiplexed illumination case~\cite{Ren.etal2017}. They
provide a favorable trade-off between the reconstruction quality and compute
time~\cite{Yeh.etal2015}. However, the very fundamental question of how one
chooses the step size has not been rigorously investigated. Since the
acquisition parameters (for instance, the illumination coding and number of
measurements) can vary, the importance of a systematic approach to determine
the step size is pronounced in practice. While there are known strategies
for quadratic cost functions of the phase retrieval
problem~\cite{Jiang.etal2016}, these approaches do not apply to other cost
functions including those that are effective for the FPM model at
hand~\cite{Yeh.etal2015}. More importantly, these heuristics do not provide us
with convergence guarantees except for idealized random models such as those in
\cite{candes2013phaselift, Candes.etal2015, soltanolkotabi2017structured}.

In this paper, we develop an auto-tuned algorithm for FPM with multiplexed
coded illumination that is based on theoretical principles. To that aim, we
propose a gradient descent algorithm called the accelerated Wirtinger Flow
(AWF). Our main contributions are: 
\begin{itemize}
\item The proposal of an analytical
expression to select the step size, which makes the final algorithm free of
tuning parameters. The framework is applicable to any type of LED selection for
multiplexing. 
\item The stationary point convergence of the Wirtinger flow
iterates with the chosen step size. 
\item The illustration of AWF's
convergence speed reaching its analogue with manually optimized step size. We
also demonstrate the applicability of AWF to real data. 
\end{itemize}
\section{Forward Model} \label{Sec:Method}
\begin{figure}[t]
  \centering
  \includegraphics[width=0.8\textwidth]{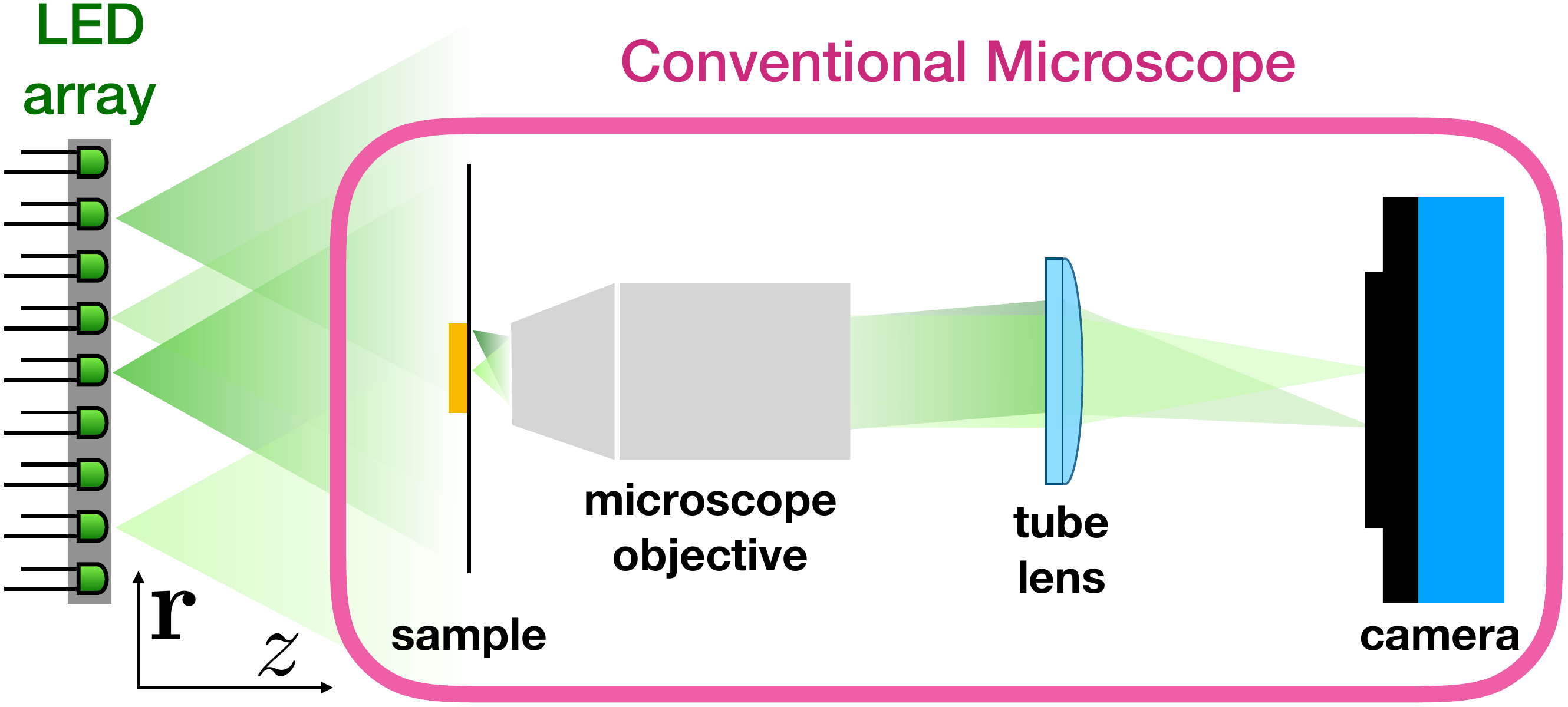}
  \caption{Optical setup and the geometry multiplexed FPM. Multiple LEDs
    simultaneously illuminate the sample from different angles. After passing
    thorugh the microscope,  sample's image is generated at the camera plane
    where intensity-only measurements are recorded.  The goal is to reconstruct
    both phase and amplitude with higher resolution than the objective's
    diffraction limit.}
  \label{fig:Setup}
\end{figure}

We start with a mathematical description of the measurement formation process
in FPM with multiplexed coded illumination. Consider the setup in
Figure~\ref{fig:Setup}, where an array of LEDs is used
as the illumination source of a standard microscope. The coordinate vector is
given by $(\rbf,z)$ where $\rbf = (r_1,r_2) \in \R^2$ denotes the spatial
location on a transverse plane perpendicular to the optical axis $z$. A
\textit{thin} sample (that is the 2D object to be imaged) is located at
$z=0$, which is also the rear focal plane of the objective lens.

The sample is characterized by a complex-valued \textit{transmission function} 
\begin{align}
  s(\rbf) \coloneqq \sqrt{I_s(\rbf)} \exp\left(
  \cj \phi_s(\rbf) \right)\hspace{-.23em}\text{,} 
  \label{eq:objectField}
\end{align}
where $\cj^2 = -1$ and the continuous mappings $I_s:\Omega\rightarrow \R$ and
$\phi_s:\Omega\rightarrow \R$ represent the spatial \textit{intensity} and
\textit{phase} maps (of the sample), respectively. The domain $\Omega$ is
assumed to be a compact subset of $\R^2$.

Placing the LED array sufficiently far away,  each LED's
illumination is modeled as a monochromatic plane wave at the sample plane
$z=0$. When the $i$th LED is switched on, the field
exiting the sample (and entering the microscope objective) is given by
$
s_{i} (\rbf) = s(\rbf) \exp \left( 2 \pi \cj \langle \xibf_{i}, \rbf
\rangle \right),
$ 
where $\xibf_{i}$ is the spatial frequency vector of the corresponding angle of
illumination\footnote{$\xibf_{i} \coloneqq (1/\lambda)( \cos(\alpha_{i}),
\cos(\beta_{i}))$ where $\lambda$ is the wavelength; $\alpha_{i}$ and
$\beta_{i}$ are the angles of incidence on the axes $r_1$ and $r_2$,
respectively.}. In effect, each $s_i$ is uniquely described by shifting the
Fourier transform\footnote{The Fourier transform $\hat{f}$ of a function $f$ is
defined by 
\begin{align*}
  \widehat{f}(\xibf) = \mathcal{F} \lbr f \rbr (\xibf) &\coloneqq \iint
  f(\rbf) \exp\left( -\cj 2 \pi \langle \xibf,\rbf \rangle \right)
  \dint\rbf\text{.}
\end{align*}
} of the transmission function $\widehat{s}$ since it holds that
$ 
  \widehat{s_{i}} (\cdot) = \widehat{s} (\cdot - \xibf_{i})\text{.} 
$

Due to the finite-aperture objective lens, the exit field is low-pass filtered
as it goes though the microscope. The process is specified by the
\textit{pupil} function $\widehat{p}$, which is the Fourier transform of the
coherent point spread function~\cite{Goodman.1996}, and suppresses spatial
frequencies beyond the diffraction limit (the pupil function is 1 inside the
numerical aperture (NA) of the objective and 0 otherwise). Consequently, the
camera captures the intensity of the lower-resolution field, expressed as:
\begin{equation}
  I_{i}(\rbf) = \Bigl \lvert \Fo^{-1} \lbr \left(\widehat{p}(\cdot) \,
  \widehat{s}(\cdot - \xibf_{i}) \right) \rbr \Bigr\lvert^2(\rbf) \text{.} 
  \label{eq:measuredImage}
\end{equation}
We see in~\eqref{eq:measuredImage} that FPM can sample spatial frequency bands
beyond the diffraction limit via angle-varying illumination, however phase
information is lost. By sampling the spatial frequency bands with overlapping
regions, FPM enables phase retrieval, but this typically requires excessive
redundancy~\cite{Zheng.etal2013}. \textit{Multiplexing} fixes this
(\textit{i.e.}~less redundancy) without making the phase retrieval fail. In
this approach, rather than turning on LEDs on one at a time, a subset of LEDs
are simultaneously lit~\cite{Tian.etal2014}. Since LEDs are mutually incoherent
with each other, the total intensity of the multiplexed measurement is
expressed as the sum of the intensity if each LED was switched on
individually:
\begin{equation}
  I_{\mathcal{M}}(\rbf) =  \sum_{i \in \mathcal{M}} I_i (\rbf)\text{,} 
  \label{eq:multiplexedImage}
\end{equation}
where $\mathcal{M}$ denotes the index set of the selected LEDs.



Finally, we discretize~\eqref{eq:multiplexedImage} by considering its
\textit{amplitude-based} counterpart as such formulations are more robust to
both noise and model mismatch~\cite{Yeh.etal2015}. To that end, let $\sbf
\in\C^{n}$ be a discretization of the Fourier transform of the transmission
function and $\ybf_{\mathcal{M}} \in \R^{m}_{\geq 0}$ the discrete samples of
$\sqrt{I_{\mathcal{M}}}$ at the camera plane. We  note that  $m < n$ since the
reconstructed transmission function has a higher space-bandwidth product than
the measured image~\cite{Zheng.etal2013}. For each multiplexed measurement, the
discretized forward model is hence given by
\begin{equation}
  \ybf_{\mathcal{M}} = \sqrt{\sum_{i \in \mathcal{M}} \,\abs{ \Fbf\cT \Pbf
  \Cibf \sbf }^2} \text{.} 
  \label{eq:discreteForwardModel}
\end{equation}
where $\Cibf \in \C^{m \times n}$ is the matrix representation of the $m$-pixel
cropping centered at $\xibf_{i}$, $\Pbf \in \C^{m \times m}$ is a diagonal
matrix generated from the discretized pupil function, and $\Fbf\cT \in \C^{m
\times m}$ represents the inverse of the discrete (normalized) Fourier transform~\cite{Bostan.etal2013}.
Note that $\abs{\cdot}^2$ and $\sqrt{\cdot}$ are element-wise operations.

\section{Reconstruction Algorithm: Accelerated \\Wirtinger Flow}
In this section we propose a reconstruction algorithm for recovering the
sample's amplitude and phase maps from a set of multiplexed intensity
measurements. To this aim, we solve the following optimization problem:
\begin{align}
  \underset{\mbf{s}\in\C^{n}}{\text{min}}\quad \left( \mathcal{J}(\mbf{s})
  \coloneqq \sum_{k=1}^K \twonorm{\ybf_{\mathcal{M}_k} - \sqrt{\sum_{i \in
  \mathcal{M}_k} \,\abs{ \Fbf\cT \Pbf \Cibf \sbf }^2} }^2 \right) \hspace{-0.23em}\text{.}  
  \label{opt}
\end{align}
Here, $K$ represents the total number of captured multiplexed images. An
iterative approach to solve this problem is to consider gradient descent type
updates of the form 
\begin{align}
\mbf{s}_{t+1} \leftarrow \mbf{s}_t-\mu_t \nabla \mathcal{J}(\mbf{s}_t),
\label{iters}
\end{align}
where $\mu_\tau$ is the step-size at iteration $t$. Since the cost function
$\mathcal{J}$ is not complex-differentiable\footnote{$\mathcal{J}$ is a mapping
from $\mathbb{C}^{n}$ to $\R_{\geq 0}$ so that it is not holomorphic.}, we
shall rely on the notion of Wirtinger derivatives to define the gradient
$\nabla \mathcal{J}$ and, hence, refer to~\eqref{iters} as Wirtinger Flow
(WF)\footnote{It is noteworthy that~\eqref{iters} is also closely related to
the well-known Gerchberg-Saxton
method~\cite{Gerchberg.Saxton1972}.}~\cite{Candes.etal2015}. Still, the cost
function is differentiable (even in the sense of Wirtinger derivatives)  except
for isolated points so that we use the notion of generalized gradients.
This allows us to define the gradient at a non-differentiable point as one of
the limit points of the gradient in a local neighborhood of the
non-differentiable point~\cite{Nocedal.Wright2006}. For our cost function in
\eqref{opt}, the generalized gradient takes the form
\begin{align}
\label{gengrad}
\nabla \mathcal{J}(\mbf{s})\coloneqq \sum_{k=1}^K
\sum_{i\in\mathcal{M}_k} \Cibf\cT \Pbf\cT \Fbf \mbf{e}\text{,}
\end{align}
where
\begin{align*}
  \mbf{e} = \left(\sqrt{\sum_{i \in \mathcal{M}_k} \,\abs{ \mbf{A}_i \mbf{s}}^2}-\ybf_{\mathcal{M}_k} \right) \odot
  \left(\frac{\mbf{A}_i \mbf{s}}{\sqrt{\sum_{i \in \mathcal{M}_k} \abs{\mbf{A}_i \mbf{s}}^2} }\right)
\end{align*}
with $\mbf{A}_i = \Fbf\cT \Pbf \Cibf$. Here, $\odot$ denotes the Hadamard (\textit{i.e.}~element-wise) 
product while $\Pbf\cT$ and $\Cibf\cT$ are the
adjoint operators of $\Pbf$ and $\Cibf$, respectively. Note
that for a complex-valued vector $\mbf{a}$, the element-wise division $\mbf{a} /
\abs{\mbf{a}}$ results in a vector whose entries contain the phase of the
entries of $\mbf{a}$. 

Based on bounded Hessian arguments\footnote{We omit the full derivation.}, we propose the following step size for the WF iterations in~\eqref{iters}:
\begin{align}
\label{step1}
\mu_t = \mu \coloneqq \frac{1}{ \twonorm{\sum_{k=1}^K\sum_{i\in\mathcal{M}_k}
\Cibf\cT \Pbf\cT \Pbf\Cibf}}.
\end{align}
This proposed step size is
general and can accommodate any illumination coding design represented by
$\mathcal{M}_k$, $k = 1,\ldots,K$. Moreover, we establish that
\begin{equation}
\twonorm{\sum_{k=1}^K\sum_{i\in\mathcal{M}_k}
\Cibf\cT \Pbf\cT \Pbf\Cibf} = \infnorm{ \sum_{i \in \mathcal{A} } \abs{\Pbf_i}^2}\text{,}
\end{equation}
where $\mathcal{A} \coloneqq \cup_{k=1}^{K} \mathcal{M}_{k}$ and
$\infnorm{\cdot}$ is the maximum norm. Here, $\Pbf_i$ represents a shifted
pupil function that is centered at $\xibf_i$. Therefore, under the assumption
of an ideal pupil function, the step size is inversely proportional to the
maximum redundancy factor of the sampling in the Fourier domain.

\subsection{Theory for convergence to stationary points}
\label{sectheory}
We now provide a theoretical justification for our choice of step
size in~\eqref{step1}. To do so, we state the following theorem:
\begin{theorem}\label{mainthm} For $k=1,2,\ldots,K$, let $\ybf_{\mathcal{M}_k}\in\mathbb{R}^{m}_{\geq0}$ denote the $k$th multiplexed FPM measurement according to~\eqref{eq:discreteForwardModel}. We run the updates 
\begin{align*}
\mbf{s}_{t+1} \leftarrow \mbf{s}_t-\mu_t \nabla\mathcal{J}(\mbf{s}_t),
\end{align*}
with $\nabla \mathcal{J}(\mbf{s})$ as defined in~\eqref{gengrad} and the step size $\mu_t$ obeying~\eqref{step1}.
Also, let $\mbf{s}_{*}\in\underset{\mbf{s}\in\C^{n}}{\arg\min} \, \mathcal{J}(\mbf{s})$ be a global optima. Then, it holds that
\begin{align*}
\underset{t \rightarrow\infty}{\lim} \twonorm{\nabla \mathcal{J}(\mbf{s}_t)}\rightarrow 0,
\end{align*}
and 
\begin{align*}
\min_{t \in\{1,2,\ldots,T\}} \twonorm{\nabla \mathcal{J}(\mbf{s}_t)}^2\le \frac{\left(\mathcal{J}(\mbf{s}_1)-\mathcal{J}(\mbf{s}_{*})\right)}{\mu T}.
\end{align*}
\end{theorem}
Our theorem demonstrates that the WF iterates with our chosen step size
converge to a point where the generalized gradient is zero. This is a
non-trivial statement, as the cost function is non-smooth and there are many
stationary points where the generalized gradient does not vanish. We note that
the theorem does not imply convergence to a local optima. 

\subsection{Acceleration} 

We have shown that the WF iterates converge to a stationary point if the step
size is set as in~\eqref{step1}. However, the convergence rate is still rather
slow. To overcome this challenge, inspired by the seminal work of
Nesterov~\cite{Nesterov.1983}, we apply an acceleration method to our
WF scheme. The accelerated WF (AWF) updates are  
\begin{subequations}
  \label{AWF}
\begin{eqnarray}
\mbf{v}_{t+1} &\leftarrow& \mbf{s}_t-\mu_t \nabla \mathcal{J}(\mbf{s}_t);  \\
q_{t+1} &\leftarrow & 1/2 + (1/2) \sqrt{1 + 4 q_t^2}; \\ 
\mbf{s}_{t+1} &\leftarrow & \mbf{v}_{t+1} + (q_t - 1/q_{t+1}) (\mbf{v}_{t+1} - \mbf{v}_{t}), 
\end{eqnarray}
\end{subequations}
where $q_1 = 1$ and the step size $\mu_t$ is kept the same. We note that
Nesterov's acceleration scheme is derived for convex and smooth functions,
which is not the case for the cost function in \eqref{opt}. Within our
framework, it is used as a remedy for improving the convergence in practice,
which we shall demonstrate in Section \ref{Sec:NumericalResults}.

\section{Numerical Results}
\label{Sec:NumericalResults}

\begin{figure}[t]
  \centering
  \includegraphics[width=0.8\textwidth]{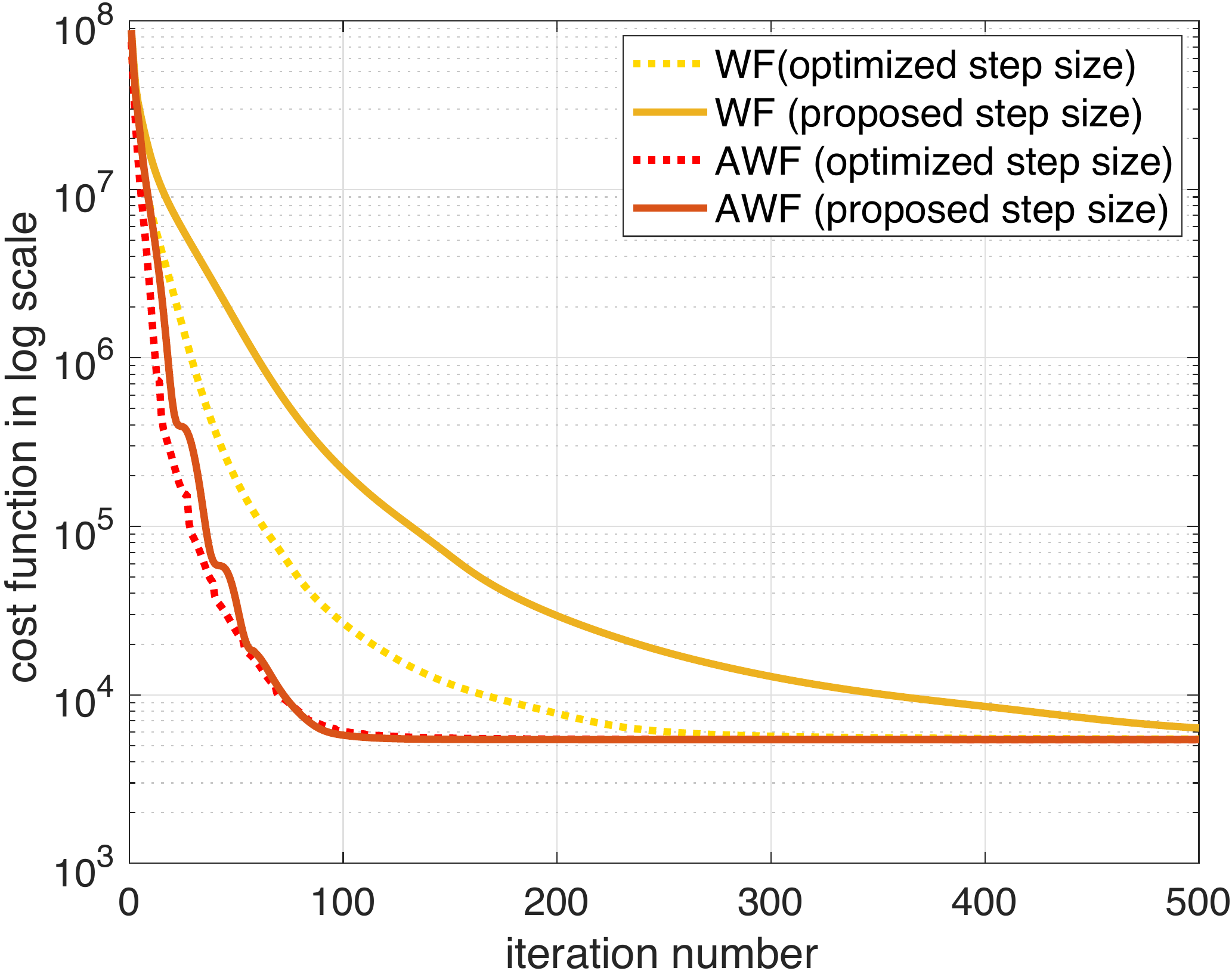}
  \caption{Evolution of the cost function throughout with iterations: Wirtinger
    Flow (WF) and its accelerated version (AWF) implement the updates
    in~\eqref{iters} and~\eqref{AWF}, respectively.}
  \label{fig:Convergence}
\end{figure}

\begin{figure}[h!]
  \centering
  \includegraphics[width=0.8\textwidth]{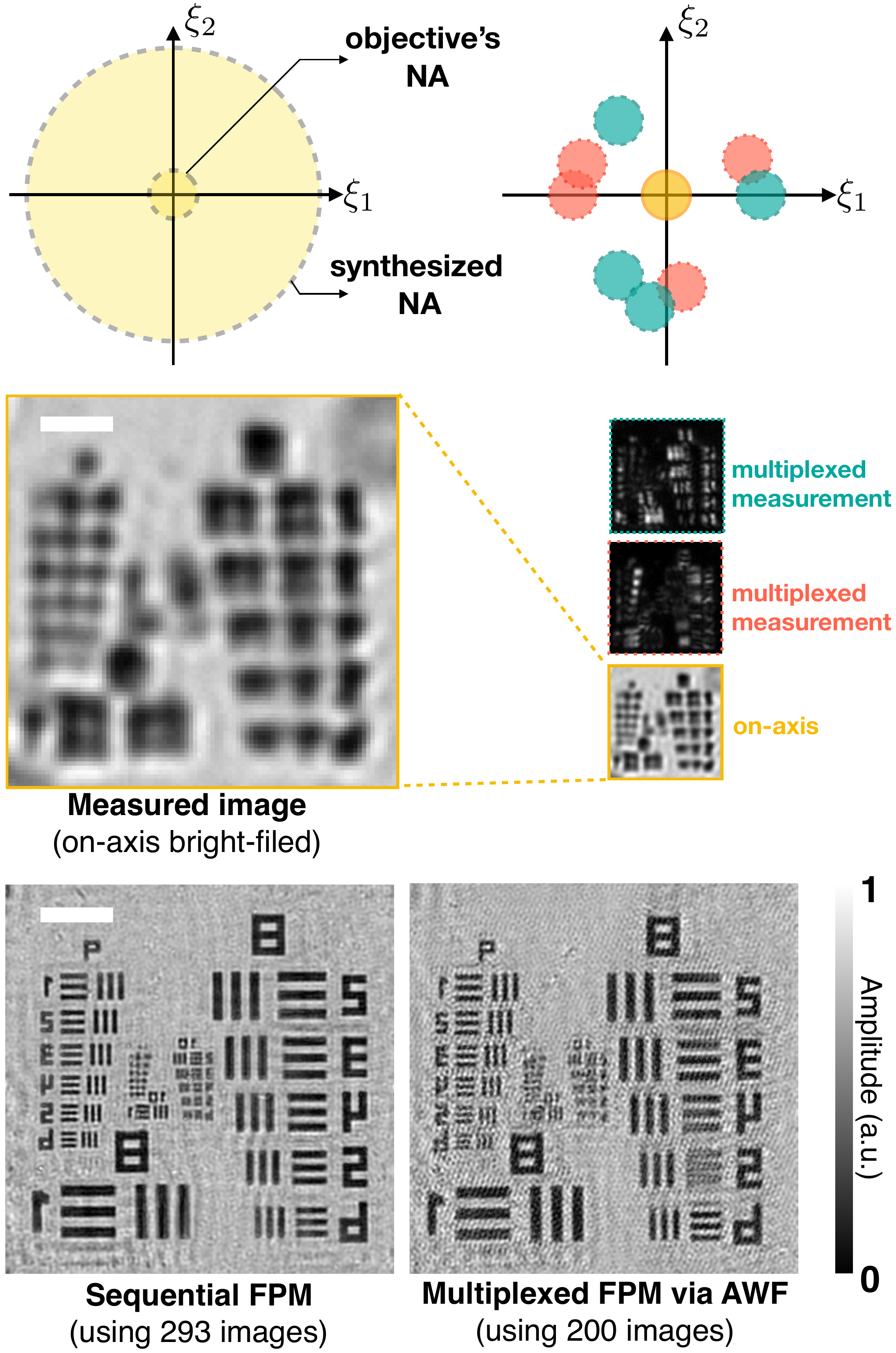}
  \caption{Application of the proposed algorithm to USAF resolution target. Scale bar is \SI{50}{\micro\metre}. See
  text for further information.}
  \label{fig:Realdata}
\end{figure}

We illustrate the practical benefits of the AWF algorithm with
both simulated and experimental data. We start by investigating
the efficiency of our analytical step size. We simulate the optical system in
Figure~\ref{fig:Setup} using the following physically-accurate parameters:
separation of the LEDs is $4$ mm;  distance of the LED array to the sample is
$77$ mm; illumination wavelength is $514$ nm; microscope objective has 0.1 NA
with $8 \times$ magnification; pixel size is \SI{6.5}{\micro\metre}. We
consider a total number of 293 LEDs. Randomly chosen 4 LEDs are lit at the same
time for multiplexing.

For comparison, we manually optimize the step size for WF where we aim at
achieving the fastest-possible convergence speed while ensuring that the cost
function decreases as the iterations proceed. Once the step size is tuned, we
also incorporate the Nesterov's acceleration. We run the algorithms for 500
iterations. All methods use the same initialization that sets a constant
image for both amplitude and phase. 

By looking at the convergence plots illustrated in
Figure~\ref{fig:Convergence}, we see that the AWF with our proposed step size
is as efficient as its variant that uses a manually-tuned step size. This shows
that our parameter-free approach does not compromise performance, providing
us with a practical framework that does not require any tuning. We see that
Nesterov's acceleration notably improves the convergence speed for both cases.
We also note that there is a significant difference between the non-accelerated
methods. This highlights the importance of acceleration strategies for
nonlinear inverse problems.

For the experimental data,  we image a USAF resolution target with a commercial
Nikon TE300 inverted microscope that uses programmable $32 \times 32$ LED
array. Experimental parameters are the same as those used in the simulations.

We consider $4 \times$ multiplexed measurements (total of 200 images) as well
as the sequential ones (total of 293 images) for validation. We use AWF with
the proposed step size for both reconstructions (see
Figure~\ref{fig:Realdata}). Our experiments show that FPM drastically improves
the spatial resolution (approximately $5 \times$) of the imaging system. We
also see that the reconstruction obtained from multiplexed measurements (with
$30\%$ less data) is similar to the one from the sequential data, but exhibits
more artifacts.


\section{Conclusion}
\label{Sec:Conclusion}

We  introduced a theoretically-sound reconsruction algorihm for multiplexed
FPM. Our main contribution has been the proposal of an analytical step size for
which we established a stationary point covergence. Considering a Nesterov-type
acceleration, we have shown that the practical convegence is as fast as the
case where the step size is manually optimized. 

\bibliographystyle{IEEEbib}
\bibliography{references}

\end{document}